\documentclass[preprint,12pt]{elsarticle}

\makeatletter
\@ifundefined{affiliation}{%
  \expandafter\def\csname affcompat@hanyang\endcsname{Department of
    Physics, Hanyang University, Seoul 04763, Republic of Korea}%
  \expandafter\def\csname affcompat@neurometry\endcsname{Neurometry,
    211 Hwarang-ro, Seongbuk-gu, Seoul 02772, Republic of Korea}%
  \newcommand{\affiliation}[2][]{%
    \address[#1]{\csname affcompat@#1\endcsname}}%
}{}
\makeatother

\usepackage{amssymb}
\usepackage{amsmath}
\usepackage{amsthm}
\usepackage{mathtools}
\usepackage{bm}
\usepackage{subcaption}
\usepackage[hidelinks]{hyperref}

\newcommand{\chivev}{\langle\chi\rangle}
\newcommand{\Fchi}{F_{\chi}}
\newcommand{\Phid}{\Phi}
\newcommand{\gw}{g_{\omega NN}}
\newcommand{\gom}{g_{\omega}}
\newcommand{\Gw}{G_{\omega}}
\newcommand{\gor}{g_{\rho}}

\newcommand{\gchi}{g_{\chi NN}}
\newcommand{\grho}{g_{\rho NN}}
\newcommand{\nB}{n}
\newcommand{\ns}{n_{s}}
\newcommand{\Tmm}{T^{\mu}_{\mu}}
\newcommand{\Tmed}{(\Tmm)_{\rm med}}

\newcommand{\pd}[2]{\frac{\partial #1}{\partial #2}}
\newcommand{\dd}[2]{\frac{d #1}{d #2}}
\newcommand{\bi}{\bibitem}

\def\be{\begin{eqnarray}}\def\ee{\end{eqnarray}}
\def\lsim{\mathrel{\rlap{\lower3pt\hbox{\hskip1pt$\sim$}}
     \raise1pt\hbox{$<$}}}
\def\le{ \begin{array}{ll}}\def\re{\end{array}}

\def\lsim{\mathrel{\rlap{\lower3pt\hbox{\hskip1pt$\sim$}}
     \raise1pt\hbox{$<$}}}
\def\gsim{\mathrel{\rlap{\lower3pt\hbox{\hskip1pt$\sim$}}
     \raise1pt\hbox{$>$}}}

  {\par\medskip}

\theoremstyle{remark}

\begin{document}

\begin{frontmatter}

\title{ Scale Invariance and Compact Star  Matter}

\author[hanyang]{Hyun Kyu Lee}
\ead{hyunkyu@hanyang.ac.kr}

\author[neurometry]{Won-Gi Paeng}
\ead{wgpaeng@neurometry.co.kr}

\affiliation[hanyang]{organization={Department of Physics, Hanyang
             University},
            city={Seoul},
            postcode={04763},
            country={Republic of Korea}}

\affiliation[neurometry]{organization={Neurometry},
            addressline={211 Hwarang-ro, Seongbuk-gu},
            city={Seoul},
            postcode={02772},
            country={Republic of Korea}}

\begin{abstract}

\noindent  We present discussions on the possibility of emerging hidden scale symmetry, as a pseudo-conformal phase in super dense baryonic matter,
using   the velocity of sound as a criterion for a scale symmetry window in hadronic dense matter.

\noindent In the density dependent mean field  approach  \`{a}  la Brown-Rho scaling,  it has been observed  that  the interplay  between vector mesons and  $ \chi$,  one of the strongly correlated  effects between hadrons,    is nontrivial  such that    the trace of the energy momentum tensor becomes density-independent in the super dense regime and the sound velocity approaches  the conformal sound velocity for the pseudo-conformal phase.
It is suggested that in the pseudo conformal phase     the rearrangement terms induced by density dependent couplings do not spoil the hidden scale symmetry in the  compact star matter.

\noindent We elaborate further on the astrophysically observable quantities of the compact stars and the implications for the parity doubling and the quark-hadron transitions.
\end{abstract}


\end{frontmatter}

\section{Introduction: From Brown-Rho scaling to  Pseudo Conformality}\label{sec1}

 In the real world, the scale transformation  is not a symmetric transformation basically because most of elementary particles possess finite masses.  Scale invariance   would be exact if all  (more generally, all  dimensionful couplings) vanish~\cite{coleman}.   But there are various  situations where the effects of the masses are not important, such as  extremely high-energy phenomena.  Then  the~role of  scale symmetry, when properly formulated,  can be studied systematically, even  together with  small  symmetry-breaking effects.
The conceptual benefit  is that even approximate realizations of scale symmetry  may reveal how scale transformations operate in the real world and how  the effects of symmetry  breaking terms  can be  analysed.   Similarly,  when we can find an appropriate conformal  window  in highly dense matter, where the density matters rather than the  kinetic momentum, we can study the~role of the  scale symmetry, when properly formulated,  systematically.

Hadronic  matter  at a much higher density  than normal nuclear density, $n_0$, has been  of great interest since  recent observations~\cite{NS, Krastep, Fonseca21, NICER21}
imply the possibility of a high-density regime, $ 5 \sim 6 \, n_0$,   at the core of  neutron stars.
 These density regimes  have not been fully explored  theoretically or experimentally and there is no idea on what kind of   new  symmetry is emerging as a result of the strong hadronic correlation.   It can be anticipated that nontrivial strong correlations of nuclear matter invoke the quest for the scale symmetry.

In 1991, Gerry Brown and Mannque Rho\cite{BR} proposed   the  formulation of  the  effective theory of hadronic matter, where the  scale symmetry  is implemented,   to discuss the sliding vacua with baryon number density.   In this scale symmetry implemented  formulation  the  predictive  power of the effective theory  is found to be transparent:    the masses of hadrons (like the nucleon $N$, and the mesons  $\rho$, $\omega$, and dilaton) scale in the nuclear medium primarily due to the partial restoration of chiral symmetry.   The idea of sliding vacua  naturally leads to the density dependent  mean field approach,  where  the coupling constants are density-dependent.    Mannque  Rho and his collaborators\cite{pklr,pklmr2}  further developed  the idea of  sliding vacua to see how   the effective theory  of  dense  nuclear matter  can reveal  the scale symmetry in   high-density regime.    One of the interesting features  of  the chiral effective  field theory\footnote{A  brief historical background  of  the effective field theory adopted here, coined as GnEFT,  can be found in \cite{rho1}.},  with the scale symmetry implemented by  the dilaton field, $\chi$, is that   the  trace of the energy--momentum tensor (TEMT)  depends only on the vacuum expectation value of the dilaton $\langle\chi\rangle$.
The vacuum expectation value of the dilaton is  basically  due to the spontaneous symmetry breakings of  scale symmetry and it turns out that the  nucleon mass at a higher density is supposed to be mainly  from the vacuum expectation value of the dilaton developed for  the spontaneously broken phase  of the scale symmetry.  The scale-symmetry-breaking effect due to nucleon mass term is hidden in $\langle\chi\rangle$.   And a non-zero value of TEMT indicates the apparent violation of the scale-invariance of the system of nuclear matter.   The equation of state of nuclear matter based on the effective theory  has been discussed by incorporating the many body effects such as  nuclear correlations  in the   renormalization group analysis, $V_{{\rm low}\,k}$, which  is  beyond the mean field   approach.

Interestingly, it  is found that     the trace of   energy momentum tensor  becomes a density-independent finite quantity at  higher density  so that     the sound velocity  approaches   that of  conformal  symmetric matter,  conformal velocity,  $v_c =1/\sqrt{3} ~c$.  It is dubbed  the pseudo-conformality of the dense hadronic matter \cite{pcspeed}.   It is the many body effect in  $V_{{\rm low}\,k}$ ,   specifically  the interplay between the omega meson excitation and nucleons  and the tensor forces dominated by $\rho$ and $\pi$ exchanges, in the dense medium,  that  reveals  the conformal velocity   for $n > n_A$. It is considered one of the characteristics of the underlying scale symmetry   in dense matter.  This feature  accounts for the emergent pseudo-conformal  symmetry in the  compact-star matter, and suggests that the core of the compact star provides a  new window  for   investigating the  effect of  the scale symmetry (spontaneously broken) hidden in the dense hadronic matter.

The  effective theory  constructed  in \cite{pklmr2}  is supposed to have the fixed point features of  Landau Fermi Liquid  at zero temperature with  relevant  coupling constants  determined  at the given density.   In this formulation, the scale symmetry breaking is solely due to the dilaton potential which develops  a  non zero vacuum expectation value  of dilaton, which determines  the    masses of  the hadrons.
 The advantage of  the dilatonic  formulation is  that  the  scale symmetry breaking  effect  represented by TEMT  relies  simply on  $\chi$.   TEMT, $T^{\mu}_{\mu}$, is given by
\begin{equation}
\label{eq:Tvacgen}
\left(T^{\mu}_{\mu}\right)_{\chi} =4\,V(\chi)-\chi\,\frac{\partial V({\chi})}{\partial \chi} .
\end{equation}
At a fixed density it seems to work. But  when we  suppose   the effective theory  should be  relevant for the wide range of  density, from normal nuclear density to star matter density,  the coupling constants should be considered to be density dependent~\cite{BR,FRS99,Song01}. These density dependences break the scale symmetry.    TEMT  gets  additional  contributions, $ (T^{\mu}_{\mu})_{\rm med}$,  from density dependent coupling constants, for example,  in the form of rearrangement terms~\cite{FLW95}.
\begin{equation}
\label{eq:Tmunu}
T^{\mu}_{\mu}=(T^{\mu}_{\mu})_{\rm \chi} +  (T^{\mu}_{\mu})_{\rm med}.
\end{equation}
This shows that  the density dependent coupling constants spoil the advantage of implementing scale symmetry   while the Fermi surface does not spoil the scale symmetry.   As a guiding conjecture  it is supposed    that the  density dependences of coupling constants should not  spoil the scale symmetry,  such that
$\left(T^{\mu}_{\mu}\right)_{\rm med}\rightarrow 0$.
We take it as a kind of consistency condition to protect the underlying scale symmetry, which controls the interplay between the dilaton and mesons under density dependent interactions.     This  interplay  in  the mean field approach  is essential  to  understand   the  emergence of the pseudo conformal phase as has been demonstrated also in the simplified mean field approach \cite{PLRS} that  $\langle\chi\rangle$  becomes density independent at high density.

In section 2,  the sound velocity of dense matter   is discussed  using  a noninteracting degenerate fermion matter at zero temperature. It is demonstrated  that a conformal window may appear in a context different from $T^{\mu}_{\mu} \rightarrow 0$, and we can take the speed of sound as a physical quantity that can quantify the conformal window, $v_s \rightarrow v_c$, even for $T^{\mu}_{\mu} \neq 0$ -- a pseudo conformal window.   The results for the equation of state and TEMT based on the effective Lagrangian with hidden local symmetry and scale symmetry implemented are reviewed in section 2.     Most of the detailed calculations and conventions are  from  the works by Paeng~et~al.~\cite{pklr,pklmr2,PLRS} and later developments~\cite{MR20, lmpr, marho2}, on the effective Lagrangian, which is  constructed with   chiral and scale symmetry,  adopted for the compact star matter in which both symmetries are spontaneously broken.
The~trace of the energy--momentum tensor  becomes  density-independent  at  higher density.    Its  implications on  the speed of sound and on  the emergence of  pseudo-conformality at higher density, are summarized.    In section 3,  the results in section 2 are discussed in the framework of  density-dependent mean field approach.  We suppose the zero temperature dense nucleon matter to be a Landau Fermi liquid with a Fermi-liquid fixed point.  We suppose
 that  the rearrangement terms induced by density dependent coupling constants and  tensor force contribution for  TEMT should be  summed to be zero not to spoil   the pseudo conformality nature.   The summary and discussions  on the observational opportunities for  the pseudo conformal phase  inside the core of the compact stars  are given in section 4.

\section{Pseudo conformal Phase}
\subsection{Degenerate Free Fermion at zero Temperature }

To discuss   whether the high density of nuclear  matter can be  a possible window  for  investigating  the  effect of the scale symmetry, we consider the trace of the energy--momentum tensor(TEMT), $ T^\mu_\mu$, which measures the scale symmetry breaking effect.     The divergence of  dilatation current, $s^{\mu}$,  induced by scale transformations, is given by
\be
\partial_{\mu} s^{\mu} =   T^\mu_\mu.
\ee
For the system with  conformal invariance,  it becomes zero,
\be
 T^\mu_\mu =0 \label{temt00}.
\ee
Then, the speed of sound $v_s$    is given by  the  conformal one,  $v_{c}= c/\sqrt{3}$.
 As an example  for the conformal window,  let us consider a noninteracting degenerate fermion matter at zero temperature with the density $n=\frac{2}{3\pi^2}\, k_F^3$ . Fermi energy and momentum are denoted  as $E_F$ and $k_F$ respectively. The  energy density and the pressure  are given by
\be
\varepsilon_{f}(m, n)
&=& \frac{1}{4\pi^2} \left[ 2E_F^3 k_F - m^{\,2}E_F k_F - m^{\,4} \ln\left( \frac{E_F + k_F}{m}\right) \right]  ,\label{efree} \\
P_{f}(m,n) &=& \frac{1}{4\pi^2} \left[ \frac{2}{3}E_F k_F^3 - m^{\,2}E_F k_F + m^{4} \ln\left( \frac{E_F + k_F}{m}\right) \right].\label{pfree}
\ee
And TEMT is given by
\be
\left(T^\mu_\mu\right)_f  &=& \varepsilon - 3P = m \,  n_s , \label{temtfree}
\ee
where $ n_s$ is the scalar density, $\langle\bar{\psi}\psi\rangle$, given by
\be
n_s \equiv  \frac{m}{\pi^2} \left[ E_F k_F - m^2\ln \left( \frac{k_F + E_F}{m}\right)\right] .
\ee
It is nonvanishing because of the   fermion mass, $m$, which breaks the scale symmetry explicitly.
At  a high density,  where  the   Fermi momentum, $k_F$,  is  much larger than the mass,
\be
k_F \gg m,
\ee
the energy density and the  pressure approach, in the leading order, to massless fermions
\be
\varepsilon \rightarrow  \frac{1}{2\pi^2} k_F^4 ,  \,\,\, P  \rightarrow \frac{1}{6\pi^2} k_F^4 = \frac{1}{3} \varepsilon.
\ee
Hence one can expect that the ultra-high density regime is a possible conformal window.
One can  naively think  that $T^\mu_\mu = \varepsilon -3 P \rightarrow 0$, but eq.(\ref{temtfree}) does not actually vanish in the high density limit. If we keep the terms  beyond leading order, in the limit $k_F \gg m$, we get
\be
\left(T^\mu_\mu\right)_f &\rightarrow&  \frac{m^2}{\pi^2}  k_F^2,  \label{thetafree}
\ee
which  is nonvanishing   and of order ${\cal O}( m^2 k_F^2)$.   On the other hand the sound velocity, $ v^2 =  c^2\frac{dP}{d \varepsilon}  $,     becomes the conformal one, $v_c$, in the limit $k_F \gg m$
\be
v^2 = v_c^2\left(1 - \frac{m^2}{k_F^2}\right) \rightarrow v^2_c . \label{vf}
\ee
It should be  noted that the speed of sound has a  conformal velocity limit, even though the trace of energy--momentum tensor does not vanish  \footnote{ It has been discussed that  a useful quantity for properly discussing the conformal limit at a high density may not be the energy--momentum tensor itself,  but   the ratio  of  the trace of  the energy--momentum tensor  to the energy density, $\Delta$,   as proposed in   \cite{delta},
\be
\Delta \equiv   \frac{T^\mu_\mu}{3 \varepsilon},
\ee
which has a proper limit for the conformal window, $k_F \gg m_N$,
\be
\Delta &\rightarrow&  {\cal O}( \frac{m^2}{ k_F^2})  \rightarrow 0 . \label{window}
\ee
The speed of sound  approaches  $v_c$ in the conformal limit, $ \Delta   \rightarrow 0$.
}.   This suggests  the sound velocity, rather than $T^\mu_\mu$,  can be used as a conformal window criterion in the dense hadronic matter. The general
expression for the speed of sound, $v_s$, is
\be
v_s^2 = c^2 \frac{dP}{d\varepsilon} =  c^2 \frac{dP}{dn}/ \frac{d\varepsilon}{dn} , \label{sound}
\ee
where $ c$ is the speed of light.  $ \frac{dP}{dn}$,   representing the compressibility of
the fermion system,   determines the restoring force, while the energy
density, $\varepsilon$, including the particle's rest
mass,  determines the inertia.    This is the speed  at which  small local fluctuations of a star matter  are  traveling.  It  depends on the equation of state and, therefore,  on the underlying hidden scale symmetry we are exploring.
The Fermi momentum of nuclear matter at a density of $6 n_0$ is $k_F \sim 0.51\, m_N$. It is not high enough for the nucleon mass to be ignored and the nucleon density $n \sim 6n_0$ cannot be a kinematic scale symmetry window in eq.(\ref{vf}).  However
 the sound velocity  depends on the density quite nontrivially compared to eq.(\ref{vf}), a criterion relevant to the non-interacting massive fermion.  The same   criterion  cannot be  simply imposed  on  the strongly interacting  fermionic matter at the core of  a compact star.    A conformal window may appear in a different context.       In fact,  it is observed that the trace of  the energy--momentum tensor appears to be constant as  a result of  many body  effects in nuclear matter, which happens  curiously  at moderately high density available in the cores of massive neutron stars. Interestingly, one can note that  the  density-independent feature  of TEMT  for the dense hadronic matter  reveals  the  conformal velocity.   It   can be taken as a  signature for the  conformal window, even though $T^\mu_\mu\neq 0$ (or $\Delta \neq 0$).   It is dubbed the pseudo-conformality\cite{pcspeed} of the  dense hadronic matter, which will be discussed below.


\subsection{ Pseudo-Conformal phase at Higher~Density}
  Nontrivial  strong correlations of nuclear matter  still invoke the quest for scale symmetry in such a high-density regime.
The relevant degrees of freedom of  dense star matter deep inside  compact stars  are   supposed  to be  hadrons  (nucleons and mesons)  appearing  in the  appropriate  effective theories, where   TEMT is a function of not only the nucleon mass and Fermi momentum but also the condensates of meson fields.
 Let us suppose that, at a higher density, the strong correlations of nucleons lead the system to reveal the hidden underlying  scale symmetry.  There would be  excitations in a scalar channel and dilaton fields,   such that  scale invariance  would be  realized formally in an effective Lagrangian with  dilatons.

  The effective field theory   which is viable  in the highly dense regime as well as in the low density nuclear matter  has been    developed in   ~\cite{pklr,pklmr2}.
 The relevant degrees of freedom  are the nucleon and the pseudo-scalar meson, pion $\pi$,  appearing in the standard chiral perturbation theory.    A dilaton field  $\chi$ is introduced  as a conformal compensator field to implement  the scale invariance  in the Lagrangian for massive matter fields.    Vector mesons, $\rho$ and $\omega$,  are introduced in the framework of  the  hidden local symmetry~\cite{HLS,hy, yamawaki}.
The effective field theory is dubbed   the dBHLS Lagrangian (coined as GnEFT) with d standing for dilaton $\chi$, B for baryon, and HLS for hidden local symmetry.  In the mean field calculation, $\langle\pi\rangle =0$. The effective Lagrangian in the  chiral limit  is  given by
\be
{\cal L}={\cal L}_{\rm inv} +{\cal L}_{\rm SB}  \label{TLag}
\ee
where
\begin{eqnarray}
{\cal L}_{\rm inv} &=& {\cal L}_N + {\cal L}_M \, \label{tLag}\\
{\cal L}_N &=&  \bar{N} \left[\, i\gamma^{\mu}\!\left(\partial_\mu - i g_{\rho NN} \vec{\rho}_\mu \cdot \vec{\tau} -i g_{\omega NN} \omega_\mu \right) - g_{\chi NN}\,\chi \,\right] N
\label{NLag}
\end{eqnarray}
\begin{eqnarray}
{\cal L}_M &=& \frac{F_{\sigma\rho}^2}{2F_\chi^2} \chi^2\, g^2_\rho\, \vec{\rho}_\mu \cdot  \vec{\rho}^{\,\mu}
+ \frac{F_{\sigma\omega}^2}{2F_\chi^2} \chi^2  g^2_\omega \omega_\mu \omega^\mu  \nonumber \\
& & - \frac{1}{4}\, \vec{\rho}_{\mu\nu} \cdot \vec{\rho}^{\,\mu\nu}
{}- \frac{1}{4} \omega_{\mu\nu}\omega^{\mu\nu}
{}+ \frac{1}{2}\partial_\mu\chi\cdot\partial^\mu\chi  \label{MLag}\\
{\cal L}_{\rm SB} &=& -V(\chi) , \label{break}
\end{eqnarray}
where $V(\chi)$ is the scale symmetry breaking potential specified below
\be
V( \chi) = \frac{m_\chi^2 F_\chi^2}{8}   \left(\frac{\chi}{F_\chi}\right)^4 \left[ \ln\left(\frac{\chi^2}{F_\chi^2}\right) - \frac{1}{2} \right] ,
\ee
It is to be noted  that TEMT is solely determined by the scale breaking potential $V(\chi)$,
\begin{equation}
\left(T^\mu_\mu\right)_{\chi}= 4V(\chi) - \chi \frac{\partial V(\chi)}{\partial \chi} = - \frac{m_{\chi}^{2}F_{\chi}^{2}}{4} \left(\frac{\chi}{F_{\chi}}\right)^{4} , \label{TEMT0}
\end{equation}
which  encodes  the trace anomaly of QCD, $\left(T^\mu_\mu\right)_{\rm QCD}=\frac{\beta(g_s)}{2g_s}G^a_{\mu\nu} G^{a\mu\nu}$.

\begin{figure}[h]
\begin{center}
\includegraphics[width=10.0cm]{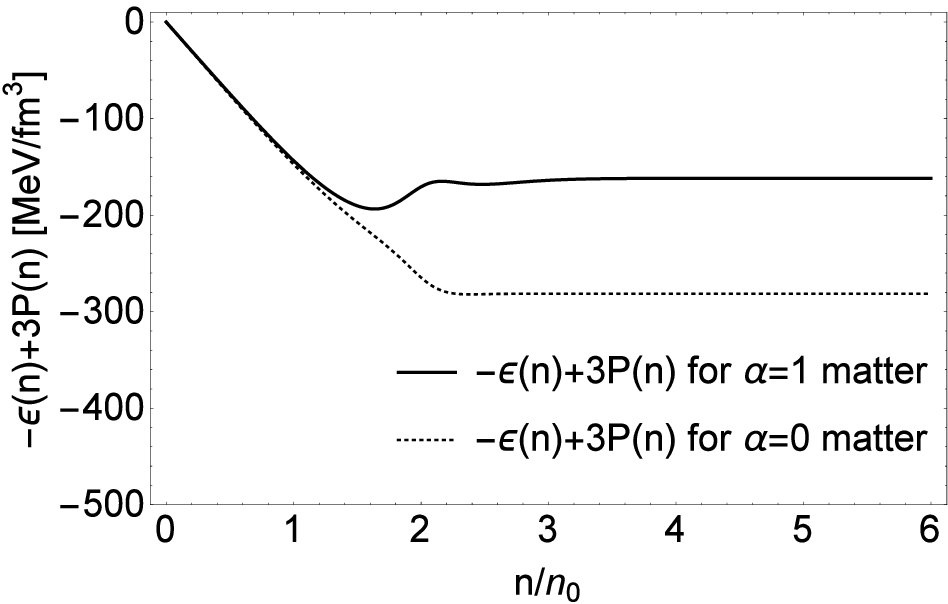}
\caption{ $-\varepsilon(n) + 3P(n) =  - \text{TEMT}$ vs. density for $\alpha=0$
(symmetric nuclear matter) and $\alpha=1$ (pure neutron matter).  } \label{TEMT}
 \end{center}
\end{figure}

 The numerical  results of the physical quantities  (for example,  symmetry energy, effective nucleon mass,  equation of  state, speed of sound, and  neutron star  properties (mass and tidal deformability)) are discussed in \cite{pklmr2} using the full  ``$V_{{\rm low}\,k}$-RG''  treatment \cite{Vlowk}.  It is observed that  TEMT becomes a density-independent finite quantity at higher density. The results are  reproduced here in  Figure \ref{TEMT} (Figure  4 in \cite{pklmr2}).
The composition of  proton number density ($n_p$) and neutron number density ($n_n$) is denoted by   $\alpha (= (n_n-n_p)/n)$, where $n(=n_n +n_p)$  is the  total nucleon number density.  Two cases for   symmetric nuclear matter $\alpha=0$  and  pure neutron matter  $\alpha=1$  are shown  among the  available compositions inside  a neutron star.  In fact, the realistic  neutron star  has a composition  between  those special cases, $0 < \alpha <1$.   The density-independence of  TEMT  at a higher density  is  expected to be valid  for the realistic neutron star matter.

It is interesting to see that,  for  the  higher  density, $n > n_A$
(here, $n_A \sim 2n_0$; the exact numerics of $n_A$  is dependent specifically on the model considered),
the numerical results of   energy density, $\varepsilon$,  and pressure, $P$,
  can be captured very well by the simple formulae
\be
\varepsilon &=&  B n^{4/3} + D  , \label{epsilon2}\\
P &=&  \frac{1}{3}B n^{4/3} - D ,
\ee
where $B$ and $D$ are density-independent parameters determined numerically \cite{pklmr2}.  Then, the trace of the energy--momentum tensor  is  simply
given by
\be
T^\mu_\mu  =  \varepsilon - 3 P = 4D, \label{temtn}
\ee
which clearly  shows the density-independence of  TEMT. As an example, for symmetric matter,  $D
= 70.4~\text{MeV/fm}^3$.

One of the interesting features of  the density plateau  of TEMT is that  the velocity of sound  turns out to be the conformal velocity $v_c$,  although the Fermi momentum corresponding to the density plateau is not high enough  to be a  kinematic conformal window in eq.(\ref{vf}).  Now the variation in the trace of the energy--momentum tensor  in eq.(\ref{temtn}) with respect to density  is zero,
\be
\frac{\partial}{\partial n} T^\mu_\mu  =  \frac{\partial}{\partial n} \left( \varepsilon - 3 P \right) =0 .
\ee
Then,
we obtain
\be
 \frac{\partial\varepsilon}{\partial n} \left(1- 3 \frac{v_s^2}{c^2}\right) =0,
\ee
and the speed of sound  becomes the conformal velocity $v_c$,
provided that there is no extremum in the energy density, $ \frac{\partial\varepsilon}{\partial n} \neq 0$.

It is the many body effect, specifically  the interplay between the omega meson excitation and nucleons in the dense medium,  that  reveals the conformal velocity for $n > n_A$~\cite{bs,dhl,bwk,AER22,BWK23b}. This is  considered one of the characteristics of the underlying scale symmetry   in dense matter.
This feature  accounts for the emergent pseudo-conformal  symmetry in the  compact-star matter, and suggests that the core of the compact star provides a  new window  for   investigating the  effect of  the scale symmetry (spontaneously broken) hidden in the dense hadronic matter.    The numerical  results of the physical quantities (for example,  symmetry energy and  neutron star  properties (mass and tidal deformability))  are  consistent  with the present  terrestrial laboratory experiments and  astrophysical observations  related to neutron stars, as discussed  in \cite{pklmr2}.

\section{Mean Field~Approach}\label{sec3}

It was discussed in  ~\cite{pklr,pklmr2,PLRS} that  the effective Lagrangian, which is  constructed with two hidden symmetries, chiral and scale symmetry,  to capture the low energy nuclear dynamics,  can be adopted for the compact star matter.  The symmetries are supposed to be realized  in  spontaneously   broken phases.  We suppose that,  in the mean field calculation,  the most relevant fields  at highly dense nuclear matter are the vector meson $\omega$,  the nucleon $N$,  and the  dilaton $\chi$ as a conformal compensator field.  We consider the effective Lagrangian of Eq.~(\ref{TLag}) in which all hadron masses
are generated by the dilaton condensate $\chivev$. Schematically, the relevant parts in Eq.~(\ref{TLag}) can be summarized
in the mean-field (Hartree) approximation appropriate to uniform matter as
\be
\label{eq:Lag}
\mathcal{L}&=&\bar\psi\!\left[i\gamma^{\mu}\partial_{\mu}
-\gchi\,\chi-\gw\,\gamma^{\mu}\omega_{\mu}
-\grho\,\gamma^{\mu}\tau_{a}\rho^{a}_{\mu}\right]\!\psi \nonumber \\
& & +\tfrac{1}{2}(\partial\chi)^{2}-V(\chi)+\mathcal{L}_{M,\rm mass},
\ee
where  the vector meson-nucleon coupling constants are $ g_{vNN} = g_v(g_{Vv}-1) $  ( $v    = \omega,\rho$).
The nucleon mass is $m_{N}=\gchi\,\chivev$, and the meson masses in $\mathcal{L}_{M,\rm mass}$ are given by
\be
m^{2}_{v } &=& g^{2}_{v }\,F_{\sigma v }^{2}\,\Phid^{2} ,
\ee
where the $\chi$ scaling (Brown--Rho scaling) is denoted by  $\Phid$,  $\Phid = \frac{\chi}{F_\chi}$.
From the thermodynamic potential~\cite{PLRS},   the energy density
\be
\varepsilon &=& \varepsilon_{f} + g_{\omega NN}\langle\omega\rangle n - \frac{1}{2} m_{\omega}^2 \langle\omega\rangle^2 \nonumber \\
& &+g_{\rho NN}\langle\rho\rangle n_3 - \frac{1}{2} m_{\rho}^2 \langle\rho\rangle^2 + V \\
&=&  \varepsilon_{f} +  \frac{n^2}{2 F_{\sigma \omega }^{2}\,\Phid^{2}} (g_{V\omega}-1) ^2  \nonumber  \\
 & & +\frac{n_3^2}{2 F_{\sigma\rho }^{2}\,\Phid^{2}} (g_{V\rho}-1) ^2   + V
\ee
where
 the gap equations for $ \rho, \omega$ variations are used,
\be
g_{\omega NN}n &=& F_{\sigma \omega }^{2}\,\Phid^{2} g_{\omega}^2 \langle\omega\rangle, \,\,  \,\, g_{\rho NN}n_3 = F_{\sigma \rho }^{2}\,\Phid^{2} g_{\rho}^2 \langle\rho\rangle . \label{gapomega}
\ee
The gap equation for $\chi$ variation is given by
\be
m_N\, n_s - F_{\sigma \omega }^{2}\,\Phid^{2} g_{\omega}^2 \langle\omega\rangle^2  - F_{\sigma \rho }^{2}\,\Phid^{2} g_{\rho}^2 \langle\rho\rangle^2 = -\left( \chi \frac{\partial{V}}{\partial{\chi}}\right)_{\langle \chi  \rangle} . \label{gapchi}
\ee

Eq.(\ref{eq:Lag}) is scale invariant except the dilaton potential  $V$ which is  the sole carrier of
explicit breaking and plays the role of the gluon condensate.  However, in medium,   the coupling constants  are supposed to acquire  intrinsic density dependences as the hadronic level imprint of the sliding QCD vacuum\cite{FRS99, Song01}
 in the spirit of Brown--Rho scaling,
\begin{equation}
\label{eq:gstar}
\gchi=\gchi(\nB),\qquad g_{vNN}=g_{vNN}(\nB),\qquad  g_v=g_v (\nB) .
\end{equation}
With density dependent coupling constants,  $g_i (n)( i =\chi, \rho, \omega) $,  there  appear  additional  scale symmetry breaking terms, known as rearrangement terms\cite{FLW95}.    It is    a kind of implicit (medium) symmetry breaking through the density
dependence \eqref{eq:gstar} of the couplings. Because the baryon density
$\nB=\langle\bar\psi\gamma^{0}\psi\rangle$ carries canonical dimension
$[\nB]=3$, any non-trivial $g(\nB)$ is irreducibly scale-non-invariant.
The genuine non-invariance is the variation of the couplings
through their argument,
\begin{equation}
\label{eq:gvar}
g_{i}(\nB)\;\longrightarrow\;
g_{i}(\nB)-3\epsilon\,\nB\,\pd{g_{i}}{\nB}+\mathcal{O}(\epsilon^{2}),
\end{equation}
which is non-zero unless $g_{i}=\text{const}$. The factor $3$ in
Eq.~\eqref{eq:gvar} is the canonical dimension of $\nB$; it is the direct
origin of the universal coefficient $-3$ that multiplies every
rearrangement contribution to the medium trace below.

Then the   chemical potential, $\mu = \partial{\varepsilon}/\partial{n}$,  in this mean-field approximation is given by
\be
\mu &=& E_F +  g_{\omega}(g_{V\omega}-1)\langle\omega\rangle  +  g_{\rho}(g_{V\rho}-1)\langle\rho\rangle\,\alpha  +\underbrace{\vphantom{\frac{1}{1}}\ns\,\chivev\,\pd{\gchi}{\nB}}_{\textstyle \mathcal{R}_{\chi}}  \nonumber \\
& &  +
\underbrace{\vphantom{\frac{1}{1}}\frac{n_3^2}{F_{\sigma \rho }^{2}\,\Phid^{2}} (g_{V\rho}-1) \frac{\partial{(g_{V\rho}-1)}}{\partial{n}}
}_{\textstyle \mathcal{R}_{\rho}} +
\underbrace{\vphantom{\frac{1}{1}}\frac{n^2}{F_{\sigma \omega }^{2}\,\Phid^{2}} (g_{V\omega}-1) \frac{\partial{(g_{V\omega}-1)}}{\partial{n}}
}_{\textstyle \mathcal{R}_{\omega}} , \label{mu}
\ee
where $n_3$  is proportional to $n$ for  fixed  proton-neutron ratio, $n_3 = \alpha ~ n$.
$\textstyle \mathcal{R}_{\chi}, ~ \textstyle \mathcal{R}_{\rho}$ and  $\textstyle \mathcal{R}_{\omega}$ are  the rearrangement terms due to the density dependences of couplings.  $\textstyle \mathcal{R}_{v}$  can be put in  different forms  using gap equations,
\be
\textstyle \mathcal{R}_{\omega} =  n\,\langle\omega\rangle\,  g_{\omega} \,  \frac{\partial{(g_{V\omega}-1)}}{\partial{n}} , \,\, \textstyle \mathcal{R}_{\rho} =  n_3\,\langle\rho\rangle\,  g_{\rho} \,  \frac{\partial{(g_{V\rho}-1)}}{\partial{n}} . \label{Rgap}
\ee

The trace of energy-momentum tensor can be calculated  by the thermodynamic relation\footnote{An alternative derivation of the medium trace  from the Lagrangian  using the Euler scaling prescription is taken up in \ref{app:flfp}.}
\be
\label{eq:T4eps}
P &=& n\mu -\varepsilon \\
\Tmm &=& \varepsilon-3P=4\varepsilon-3\mu\,\nB
\ee
Using eqs.(\ref{mu}) and (\ref{Rgap}) (via the
free-Fermi-gas identity $4\,\varepsilon_f-3E_{F}\nB=m_{N}\ns$),
 contributions  from rearrangement terms  can be calculated as given by
\be
\label{eq:medthermo}
\Tmed &=&-3\,\nB\,(\mathcal{R}_{\chi}+\mathcal{R}_{\rho} +\mathcal{R}_{\omega}) \\
&=& -3\,\ns\chivev\!\left(\nB\pd{\gchi}{\nB}\right) \nonumber \\
& & -3\,\nB_3 \, g_{\rho} \langle\rho\rangle\!\left(\nB\pd{(g_{V\rho}-1)}{\nB}\right)
-3\,\nB\,\gom\langle\omega\rangle\!\left(\nB\pd{(g_{V\omega}-1)}{\nB}\right) . \label{Tmed1}
\ee
Combining eqs.~(\ref{TEMT0}) and (\ref{Tmed1}), the complete trace is
\be
\label{eq:Tfull}
\Tmm &=& \left(T^{\mu}_{\mu}\right)_{\rm \chi} +  \Tmed \\
&=&   4\,V(\chi)-\chi\, \frac{\partial{V(\chi)}}{\partial{\chi}}   -3\,\ns\chivev\!\left(\nB\pd{\gchi}{\nB}\right) \nonumber \\
& & -3\,\nB_3 \, g_{\rho}\langle\rho\rangle\!\left(\nB\pd{G_\rho}{\nB}\right)
-3\,\nB\,\gom\langle\omega\rangle\!\left(\nB\pd{G_{\omega}}{\nB}\right),
\ee
where $G_v = g_{Vv}-1$.
At zero density the medium part vanishes -- every term carries an explicit
factor $\nB\,\partial_{\nB}g \to0$ (the running of $\Gw$ included) -- but the $\chi$ part  does not: as
$\chivev\to\Fchi$ ($\Phid\to1$), $\left(T^{\mu}_{\mu}\right)_{\rm \chi}\to-m_{\chi}^{2}\Fchi^{2}/4$. Hence
\begin{equation}
\label{eq:zerolimit}
\lim_{\nB\to0}\Tmm=-\frac{m_{\chi}^{2}\Fchi^{2}}{4}\neq0 ,
\end{equation}
the hadronic image of the (non-zero) QCD vacuum trace anomaly
$\langle(\beta(g_s)/2g_{s})G^{2}\rangle$.

The trace of the energy-momentum tensor is

\begin{equation}
\label{eq:trace}
T^{\mu}_{\mu} = - \frac{m_\chi^2 F_\chi^2}{4} \left(\frac{\langle \chi \rangle}{F_\chi}\right)^4 + (T^{\mu}_{\mu})_{\text{med}} ,
\end{equation}
where the density-dependent  couplings induce an anomalous contribution on top of the dilaton potential contribution.   Here  we propose  a working  ansatz that  the anomalous
 contribution  approaches zero at higher density
\begin{equation}
 (T^{\mu}_{\mu})_{\text{med}}  \rightarrow 0 .
\end{equation}
We consider it as a consistency condition for the scale symmetric construction which might be  protected  from the anomalous contributions incurred due to density dependent coupling constants.

\begin{figure}[htbp]
\centering
\includegraphics[width=0.78\linewidth]{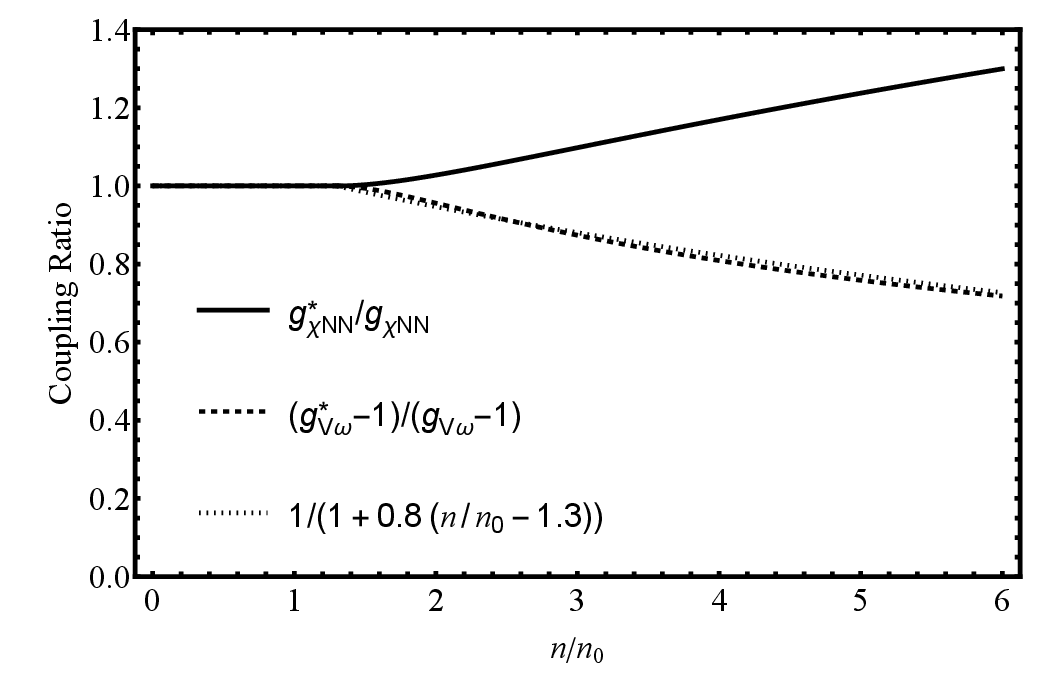}
\caption{Density running of the effective couplings entering the medium trace,
normalized to their vacuum values.}
\label{fig:coupling}
\end{figure}

\begin{figure}[htbp]
\centering
\includegraphics[width=0.78\linewidth]{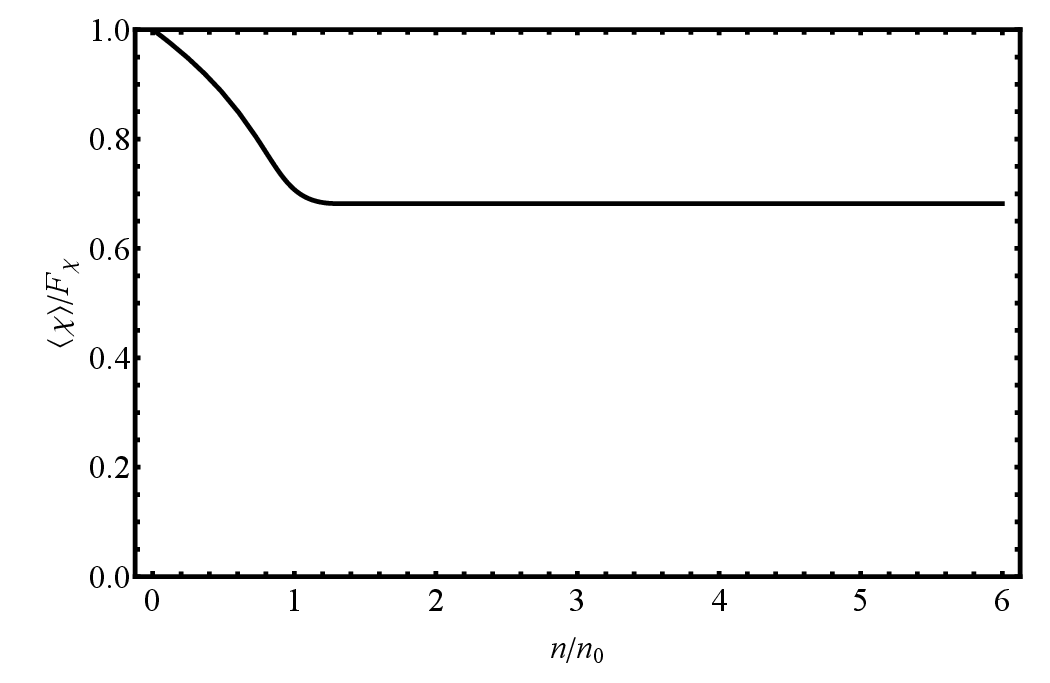}
\caption{Dilaton condensate $\Phid=\chivev/\Fchi$ as a function of density. It
departs from its vacuum value $\Phid(0)=1$ and freezes onto a plateau
$\Phid\to\Phid_{0}$. On the plateau the medium trace vanishes,
$(\Tmm)_{\rm med}\to0$.}
\label{fig:chi}
\end{figure}

To see how the consistency condition is working, we consider  the   symmetric nuclear matter, where  $n_{3}=0$ and the isovector channel drops out,
leaving the scalar (dilaton) and vector ($\omega$) runnings. Restoring the
explicit factors and using $\omega_{0}\equiv\langle\omega\rangle$, this is
\begin{equation}
\label{eq:medfinal}
\Tmed=
-3\,\ns\,\chivev\!\left(\nB\,\pd{\gchi}{\nB}\right)
-3\,\nB\,\gom\langle\omega\rangle\!\left(\nB\,\pd{G_{\omega}}{\nB}\right),
\end{equation}
which is the medium trace of symmetric matter in closed form.  In realistic parametrizations $\partial_{\nB}\gw<0$ while
$\partial_{\nB}\gchi>0$ for $\nB>\nB_{c}$,
they carry opposite  signs and can cancel.
We can  obtain the appropriate density dependencies of coupling constants, $(g_{V\omega}^{*}-1)$ and $g_{\chi NN}^{*}$ depicted in Fig.~\ref{fig:coupling}, which satisfy the consistency condition $\Tmed \rightarrow 0$.  We find that the omega-nucleon coupling exhibits a density dependence similar to the results reported in \cite{pklmr2}. We simultaneously solve  the dilaton gap equation eq.~(\ref{gapchi})    to    obtain a density dependence of  $\chi$ as shown in Fig.~\ref{fig:chi}  confirming the previous result \cite{PLRS}.
Here, we assume $n_{c} = 1.3n_{0}$, and the undetermined parameters
$g_{\omega NN}\simeq 7.2$ and $m_{\chi}^{2}F_{\chi}^{2}\simeq(279~\text{MeV})^{4}$
are determined by fitting the empirical binding energy of symmetric nuclear matter at saturation density.

Within the present scale-invariant HLS$+$dilaton framework this consistency condition is indeed realized. Solving the in-medium gap equations, the dilaton
condensate decreases from its vacuum value and \emph{saturates} onto a plateau
$\Phid\to\Phid_{0}$ (Fig.~\ref{fig:chi}). On this plateau $(\Tmm)_{\rm med}\to0$.
Then the energy-momentum tensor
is solely determined by the vacuum condensate of the dilaton field, which becomes constant  at higher density.
Hence TEMT  becomes constant.
One can see that this mean-field (density-dependent) result is qualitatively consistent with the $V_{{\rm low}\,k}$ calculation, as shown in Fig.~\ref{TEMT}. It demonstrates clearly the interplay between the $\omega$ meson and the dilaton $\chi$.

On the plateau
TEMT is given by
\begin{equation}
\label{eq:pc}
\Tmm\;\xrightarrow[\ \dd{\Phid}{\nB}\to0\ ]{}\;
-\frac{m_{\chi}^{2}\Fchi^{2}}{4}\,\Phid_{0}^{4}=\text{const}\neq0,
\qquad v_{s}^{2}\to\tfrac13 .
\end{equation}
The sound speed reaches the conformal value while the trace itself remains
non-zero, frozen at the residual dilaton-potential (gluon-condensate)
anomaly.  It is considered to be a conformal window,
where $\Tmm=\varepsilon-3P$ is constant.  Differentiating it with respect
to $\nB$ gives $d\varepsilon/d\nB=3\,dP/d\nB$, i.e.
\begin{equation}
\label{eq:vs13}
v_{s}^{2}=\frac{dP}{d\varepsilon}=\frac13 .
\end{equation}
So the conformal sound speed is realized, even though full conformality $\varepsilon=3P$ has not yet been reached. This is an emergent
pseudo-conformal symmetry: scale invariance is realized in the medium up to
the  dilaton potential, which alone survives in $\Tmm$. It is the
in-medium counterpart of the chiral--scale picture in which an infrared fixed
point coexists with explicit breaking by the dilaton mass~\cite{MR20,CT15}.

In this mean field approach, various physical effects included in the $V_{{\rm low}\,k}$ calculation presented in \cite{pklmr2} are not treated properly. For example, the tensor force has not been discussed in this mean field calculation. It is necessary to develop a proper way to treat these effects beyond the simple mean field approximation in a realistic mean field framework.     The expressions above retain only the $\omega$ and dilaton channels because, in
isospin-symmetric matter at the Hartree (mean-field) level, the $\rho$ meson
does not contribute: isospin symmetry enforces $\langle\rho\rangle=0$, so its
mean field drops out of Eq.~\eqref{eq:Tfull}. However,
  together with the pion, the $\rho$ enters
through the tensor force, whose Fock (exchange) contributions are absent
from the Hartree trace; these $\rho$- and $\pi$-mediated tensor terms add
rearrangement terms in
Eq.~\eqref{eq:medthermo} and so modify the $\omega$+dilaton-only form of $\Tmed$.
Because the tensor force is precisely the part of the nuclear interaction that
the renormalization-group evolution of realistic potentials reshapes most
strongly, it is natural to expect that, once it is included, the density
dependence of the effective couplings is driven toward the form obtained in
$V_{{\rm low}\,k}$ (low-momentum interaction) calculations~\cite{Vlowk}.

\section{Summary and Discussion}\label{sec4}

The possibility of  revealing the hidden  scale symmetry in a dense baryonic matter is discussed  in  the
framework of the effective theory for baryons and mesons where  an idea of ``sliding-vacuum'' structure of baryonic (Brown--Rho scaling) matter captured by the vacuum condensate of  dilaton $\chi$ is implemented.  It is found that
 the trace of  the energy--momentum tensor  becomes  density-independent at higher density, and the speed of sound approaches  the conformal velocity  of  the scale symmetric matter.  It is not truly conformal, $T^{\mu}_{\mu} \neq 0$, and the truly conformal regime is highly unlikely to be reached at the center of compact stars. This is the reason why it is dubbed pseudo conformal for compact stars with central density of order $6 n_0$.    It is  interpreted as an  indication that the hidden scale symmetry is emerging in high density compact star matter, disguised in the  form of the  conformal speed of sound.
 This is due to the nontrivial interplay between the $\omega$-nucleon and dilaton ($\chi$)-nucleon couplings, encoded in the asymptotic behavior of the $\omega$ coupling constant $g_{V\omega}$~\cite{pcspeed}. This feature  accounts for the emergent pseudo-conformal phase in compact-star matter, and suggests that the core of the compact star provides a  new window  for   investigating the  effect of scale symmetry (spontaneously broken) hidden in dense medium.
There are a  number of papers dealing with the sound velocity theoretically, using various models confronted with extensive analyses of experimental observations~\cite{bs,dhl,bwk,AER22,BWK23b,MMRS23,Annala23,srr}.  At present there are no serious conflicts with the available
observables, as far as  masses and radii are concerned theoretically or observationally.    Recently, the gravitational-wave observation of GW170817~\cite{GW170817} has added the tidal deformability  to the list of observables for compact stars.    Most neutron stars are populated in the mass range $1.3$--$1.5\, M_{\odot}$.  In this  mass range    the core density is not higher than  $2 n_0$.   Hence the   tidal deformability inferred from GW170817  for  $1.4 M_{\odot}$   may not be  a direct test of the emergence of the  pseudo conformal phase.   For the higher mass range $M \gsim 2\,M_{\odot}$ the inferred central densities  are  in the range  $2.5 \sim 5.0 \, n_0$,  high enough for the effect of the pseudo conformal  phase to be analysed.   Future improvements of the advanced LIGO sensitivity and next-generation gravitational-wave detectors are expected to measure the tidal deformabilities of high-mass neutron stars, clarifying the pseudo conformality in compact stars -- the simple structure depicted in Fig.~\ref{TEMT}.

In a density-dependent mean field approach,  there are two types of the scale symmetry breaking in the effective Lagrangian.   One of them is the dilaton potential as a hadronic image of  anomalous  scale symmetry breaking of QCD: QCD trace anomaly represented by  non vanishing  trace of energy momentum tensor $\left(T^{\mu}_{\mu}\right)_{\chi}$.  The density dependent coupling constants break scale symmetry implicitly and TEMT gets an additional contribution   $\left(T^{\mu}_{\mu}\right)_{\rm med}$ in the form of rearrangement terms, which  spoils the advantage of implementing scale symmetry in the effective Lagrangian, while  the Fermi surface of massive nucleon does not spoil the scale symmetry.   In this work as a guiding principle  it is conjectured    that the  density dependences of coupling constants should not  spoil the scale symmetry,
$\left(T^{\mu}_{\mu}\right)_{\rm med}\rightarrow 0$.
We take it as a kind of consistency condition to protect the underlying scale symmetry at least in higher density. It controls the interplay between the dilaton and mesons under density dependent interactions and leads to the density independent plateau of TEMT as the emergence of the pseudo conformal phase.

The  nucleon mass at a higher density is supposed to be mainly  from the vacuum expectation value of the dilaton, $\langle\chi\rangle$, developed in  the spontaneously broken phase  of the scale symmetry.  The finite nucleon mass which is unconnected to chiral condensation  is  apparently chiral-symmetry-breaking.   There should be  a nontrivial interplay between the scale and chiral symmetry   to make the  system  chiral invariant, which is considered to be parity doubling in the compact star matter.   It can be understood that  the constraint  of the chiral  symmetry  induces  the parity doubling in the pseudo-conformal matter.
 The parity doubling is an emergent phenomenon   in the pseudo-conformal phase of  the strongly correlated  compact star matter~\cite{MR20,lmpr}.   The precise mechanism by which particle excitations arrange themselves such that the system becomes pseudo-conformal with a parity-doubled structure in the nucleon sector is not yet clear, and how the populations of $N_+$ and $N_-$ in the parity-doubled structure evolve with density in the pseudo-conformal phase has not been properly discussed, while related aspects have been discussed in various contexts either in the PDM \cite{mgkh, dsz, jp} or in the pseudo-conformality~\cite{pklr, pklmr2}.
These can be interesting future directions toward the equation of state, providing useful hints  for  exploring the quark--hadron continuity in dense compact star  matter~\cite{fujimoto2,rho23}.

\vspace{6pt}

This work is a tribute to Mannque Rho, dedicated with deep respect and gratitude to our friend, great mentor, and close collaborator. Most of this work is based on discussions and ideas developed in the World-Class University (WCU) Project {\em Hadronic Matter under Extreme Conditions} (supported by the National Research Foundation of Korea, 2008--2013) under the direction of Mannque Rho as a Distinguished Professor of Hanyang University.

{\em Won-Gi Paeng}: I first met Professor Rho in 2009, when I entered
the Ph.D. program in the first year of the WCU Project, and that encounter
set the course of my academic career. He taught me what nuclear physics
truly is, and showed me that it was precisely the subject I had been
searching for. His guidance and encouragement have never ceased -- even
after I moved into the field of artificial intelligence,
he wrote to me to ``come up with an AI solution to what came
out of our 5-year WCU/Hanyang Program where you played the role of making
the crucial contribution to what is now referred to as -- lacking a more
imaginative title -- `Pseudo-conformal structure of Superdense
Matter'\,'' -- ``How about getting a beautiful AI solution to this `ugly
duckling'? It would be just wonderful!''
For this profound kindness I wish to express my deepest and most
respectful gratitude.

\appendix
\section{Euler-scaling derivation of the medium trace}
\label{app:flfp}
It is to be noted that an alternative derivation of the medium trace is possible from the Lagrangian using the Euler scaling prescription,
\begin{equation}
\label{eq:euler1}
\Tmed= 3 \,\nB\sum_{i}\pd{g_{i}}{\nB}\,
\pd{\mathcal{L}}{g_{i}},
\qquad g_{i}\in\{\gchi,\,\gw,\,\grho,\,\gom,\,\gor\} .
\end{equation}
The factor $3$ in the above equation  is the canonical dimension of $\nB$.  Using the partial derivatives
\begin{equation}
\label{eq:dLint}
\begin{gathered}
\pd{\mathcal{L}_{\rm int}}{\gchi}=-\,\chi\,\ns,\qquad
\pd{\mathcal{L}_{\rm int}}{\gw}=-\,\omega_{0}\,\nB, \qquad \pd{\mathcal{L}_{\rm int}}{\grho}=-\,\rho_{0}\,n_{3},  \\
\qquad
\pd{\mathcal{L}_{M,\rm mass}}{g_{\rho}}=+\,\frac{m^2_{\rho}}{g_{\rho}}\,
\rho_{0}^{2}, \,\,\,\, \qquad   \pd{\mathcal{L}_{M,\rm mass}}{g_{\omega}}=+\,\frac{m^2_{\omega}}{g_{\omega}}\,
\omega_{0}^{2},
\end{gathered}
\end{equation}
one can get eq.(\ref{Tmed1}), where $\langle\rho\rangle = \rho_0$ and $\langle\omega\rangle = \omega_0$: upon using the gap equations (\ref{gapomega}), the contributions of the intrinsic running of $g_{v}$ cancel between the vertex terms (through $g_{vNN}=g_{v}(g_{Vv}-1)$) and the mass terms, leaving only the running of $G_{v}=g_{Vv}-1$ in eq.(\ref{Tmed1}).


\begin{thebibliography}{999}
\bi{coleman}Coleman, S.  \textit{ Aspects of Symmetry: Selected Erice lectures}, Cambridge University Press, Cambridge, U.K., 1985.

\bi{NS}   Lattimer, J.M.   Neutron stars and the nuclear matter
equation of state. {\em  Ann. Rev. Nucl. Part. Sci.} {\bf 2021}, {\em 71}, 433.

\bi{Krastep}  Krastev, P.G.   A deep learning approach to extracting nuclear matter properties from neutron star
observations. {\em  Symmetry} {\bf 2023},  {\em 15}, 1123.

\bi{Fonseca21} Fonseca, E. \emph{et al.}  Refined mass and geometric measurements of the high-mass PSR J0740+6620. {\em Astrophys. J. Lett.} {\bf 2021}, {\em 915}, L12.

\bi{NICER21} Miller, M.C. \emph{et al.}  The radius of PSR J0740+6620 from NICER and XMM-Newton data. {\em Astrophys. J. Lett.} {\bf 2021}, {\em 918}, L28;
Riley, T.E. \emph{et al.}  A NICER view of the massive pulsar PSR J0740+6620 informed by radio timing and XMM-Newton spectroscopy. {\em Astrophys. J. Lett.} {\bf 2021}, {\em 918}, L27.

\bi{BR} G. Brown and M. Rho,  “Scaling effective Lagrangians in a dense medium,” {\em Phys.
Rev. Lett.} {\em 66}, 2720 (1991).


\bi{pklr}  Paeng, W.G.; Kuo, T.T.S.;  Lee, H.K. ;    Rho, M.
  Scale-invariant hidden local symmetry, topology change and dense baryonic matter
 {\em Phys. Rev. C} {\bf 2016}, {\em 93}, 055203.


\bi{pklmr2}
 Paeng, W.G.; Kuo, T.T.S.;  Lee, H.K. ;  Ma, Y.L.;   Rho, M.
  Scale-invariant hidden local symmetry, topology change and dense baryonic matter II.
  {\em Phys. Rev. D} {\bf 2017}, {\em 96}, 014031.

\bi{rho1}  M. Rho, A Bottom-Up EFT Approach To Superdense Baryonic Matter, {\em arXiv}, arXiv:2511.21141 (2025).





\bi{pcspeed} Rho, M. Pseudo-conformal sound speed in the
core of compact stars. {\em  Symmetry} {\bf 2022}, {\em 14}, 2154.



\bibitem{FRS99} B.~Friman, M.~Rho, and C.~Song,
``Scaling of chiral Lagrangians and Landau Fermi liquid theory for dense
hadronic matter,'' Phys.\ Rev.\ C \textbf{59}, 3357 (1999).

\bibitem{Song01} C.~Song,
``Dense nuclear matter: Landau Fermi-liquid theory and chiral Lagrangian with
scaling,'' Phys.\ Rep.\ \textbf{347}, 289 (2001).


\bibitem{FLW95} C.~Fuchs, H.~Lenske, and H.~H.~Wolter,
``Density-dependent hadron field theory,''
Phys.\ Rev.\ C \textbf{52}, 3043 (1995).


\bi{PLRS}
 Paeng, W.G. ; Lee, H.K.   ; Rho, M. ; Sasaki, C.
 Interplay   between   omega-Nucleon   Interaction and   Nucleon   Mass   in   Dense  Baryonic   Matter .
 {\em  Phys. Rev.  D} {\bf    2013}, {\em 88}, 105019 .


\bibitem{MR20} Y.-L.~Ma and M.~Rho,
``Towards the hadron--quark continuity via a topology change in compact
stars,'' Prog.\ Part.\ Nucl.\ Phys.\ \textbf{113}, 103791 (2020).


\bi{lmpr} Lee, H.K. ;    Ma, Y.L. ; Paeng, W.G. ; Rho, M.  Cusp in the symmetry energy, speed of sound in neutron stars and emergent pseudo-conformal symmetry. {\em Modern Physics Letters A}, {\bf 2022}, {\em 37}, 2230003.



\bi{marho2} Rho, M.  ; Ma, Y.L.   Manifestation of
hidden symmetries in baryonic matter: From finite nuclei
to neutron stars. {\em  Mod. Phys. Lett. A} {\bf 2021}, {\em 36},  2130012.


\bibitem{delta} Y.~Fujimoto, K.~Fukushima, L.~D.~McLerran, and
M.~Prasza\l owicz, ``Trace anomaly as a signature of conformality in neutron
stars,'' Phys.\ Rev.\ Lett.\ \textbf{129}, 252702 (2022).



\bi{HLS}  Bando, M.;  Kugo,  T. ;  Uehara,  S.; Yamawaki  K. ;
 Yanagida, T.  Is rho meson a dynamical gauge boson of
hidden local symmetry? {\em  Phys. Rev. Lett.} {\bf 1985}, {\em  54}, 1215.

\bi{hy}Harada,  M.  ;  Yamawaki, K.  Hidden local symmetry at
loop: A New perspective of composite gauge boson and
chiral phase transition. {\em Phys. Rept.} {\bf 2003}, {\em 381}, 1.

\bi{yamawaki}Yamawaki,  K.  Proving  rho meson be a dynamical gauge
boson of hidden local symmetry. {\em  Symmetry} {\bf 2023}, {\em  15}, 2209.

\bibitem{Vlowk} S.~K.~Bogner, T.~T.~S.~Kuo, and A.~Schwenk,
``Model-independent low momentum nucleon interaction from phase shift
equivalence,'' Phys.\ Rep.\ \textbf{386}, 1 (2003)







\bi{bs}  Bedaque, P. ; Steiner, A. W. ,  Sound velocity bound and neutron stars, {\em Phys. Rev. Lett.} {\bf   2015},  {\em 114}, 031103.

\bi{dhl}  Drischler, C. ; Han, S. ;  Lattimer, J. M.; Prakash, M.; Reddy, S.; Zhao, T., Limiting masses and radii of neutron stars and their implications, {\em Phys. Rev. C} {\bf 2021}, {\em 103}, 045808.

\bi{bwk} Brandes, L.   ; Weise, W.   ; Kaiser, N., Inference of the sound speed and related properties of neutron stars, {\em Phys. Rev. D} {\bf 2023}, {\em 107}, 014011.


\bi{AER22} Altiparmak, S. ; Ecker, C. ; Rezzolla, L.  On the sound speed in neutron stars. {\em Astrophys. J. Lett.} {\bf 2022}, {\em 939}, L34.

\bi{BWK23b} Brandes, L. ; Weise, W. ; Kaiser, N.  Evidence against a strong first-order phase transition in neutron star cores: Impact of new data. {\em Phys. Rev. D} {\bf 2023}, {\em 108}, 094014.





\bibitem{CT15} R.~J.~Crewther and L.~C.~Tunstall,
``$\Delta S=1$ effective chiral Lagrangian and an infrared fixed point,''
Phys.\ Rev.\ D \textbf{91}, 034016 (2015).




\bi{MMRS23} Marczenko, M. ; McLerran, L. ; Redlich, K. ; Sasaki, C.  Reaching percolation and conformal limits in neutron stars. {\em Phys. Rev. C} {\bf 2023}, {\em 107}, 025802.

\bi{Annala23} Annala, E. ; Gorda, T. ; Hirvonen, J. ; Komoltsev, O. ; Kurkela, A. ; N\"attil\"a, J. ; Vuorinen, A.  Strongly interacting matter exhibits deconfined behavior in massive neutron stars. {\em Nat. Commun.} {\bf 2023}, {\em 14}, 8451.

\bi{srr} Stroud, J.; Radice, D.; Reddy, S.  High-density sound speed and post-merger dynamics. {\em arXiv} {\bf 2026}, arXiv:2607.15588.


\bibitem{GW170817} B.~P.~Abbott \emph{et al.}
(LIGO Scientific and Virgo Collaborations),
``GW170817: Observation of gravitational waves from a binary neutron star
inspiral,'' Phys.\ Rev.\ Lett.\ \textbf{119}, 161101 (2017).




\bi{mgkh} Minamikawa, T. ; Gao, B. ; Kojo, T. ; Harada, M.  Chiral Restoration of Nucleons in Neutron Star Matter: Studies Based on a Parity Doublet Model. {\em Symmetry} {\bf 2023}, {\em 15}, 745.

\bi{dsz}    Dexheimer, V;  Schramm, S.;  Zschiesche, D.  Nuclear Matter and Neutron stars in a Parity Doublet Model. {\em Phys. Rev. C} {\bf 2008}, {\em 77}, 025803.

\bi{jp} Eser, J. ; Blaizot, J.-P.  Thermodynamics of the parity-doublet model: Symmetric nuclear matter and the chiral transition. {\em Phys. Rev. C} {\bf 2024}, {\em 109}, 045201.



\bi{fujimoto2}Fujimoto, Y.; Kojo, T.; McLerran, L.D. Momentum Shell in Quarkyonic Matter from Explicit Duality:
A Dual Model for Cold, Dense QCD.  {\em Phys. Rev. Lett.} {\bf 2024} , {\em 132}, 112701.






\bi{rho23}Rho, M.  Dense baryonic matter predicted in ``Pseudo Conformal Model''. {\em  Symmetry} {\bf 2023}, {\em 15}, 1271.








\end{thebibliography}
\end{document}